\documentstyle[12pt,aasms4,amssym,psfig]{article}

\slugcomment{To be submitted to Ap.J. Letters}

\begin{document}

\title{Discovery of Correlated Behavior Between the HXR and the Radio Bands in Cygnus X-3}

\author{M. L. McCollough, C. R. Robinson, S. N. Zhang}
\affil{Universities Space Research Association, Huntsville, AL 35805, U.S.A.}
\author{B. A. Harmon}
\affil{NASA/Marshall Space Flight Center, Huntsville, AL 35812, U.S.A.}
\author{R. M. Hjellming}
\affil{National Radio Astronomy Observatory, Socorro, NM 87801, U.S.A.}
\author{E. B. Waltman, R. S. Foster} 
\affil{Naval Research Laboratory, Washington, DC 20375, U.S.A.} 
\author{F. D. Ghigo}
\affil{National Radio Astronomy Observatory, Green Bank, WV 24944, U.S.A.}
\author{M. S. Briggs, G. N. Pendleton}
\affil{Department of Physics, University of Alabama in Huntsville, Huntsville, AL 35899, U.S.A.}
\and
\author{K. J. Johnston}
\affil{United States Naval Observatory, Washington, DC 20392, U.S.A.}

\begin{abstract}

          Using CGRO/BATSE hard X-ray (HXR) data and GHz radio monitoring data from 
the Green Bank Interferometer (GBI), we have performed a long term study ($\sim$  
1800 days) of the unusual X-ray binary Cyg X-3 resulting in the discovery of 
a remarkable relationship between 
these two wavelength bands.  We find that, during quiescent radio states, the radio flux is strongly
anticorrelated with the intensity of the HXR emission.   The relationship switches to a 
correlation with the onset of major radio flaring 
activity.  
During major radio flaring activity the HXR drops to a very low intensity during quenching in the radio and
recovers during the radio flare.  
Injection of plasma into the
radio jets of Cyg X-3 occurs during changes in the HXR emission and
suggests that disk-related and jet-related components are responsible for the high energy emission.

\end{abstract}

\keywords{ X-ray: stars; radio continuum: stars; Stars: individual: Cygnus X-3}

\section{INTRODUCTION}

     Cyg X-3 is a very unusual X-ray binary  
(see \cite{boncha} for a review) which does not 
fit well into any of the established classes of X-ray binaries.  
A 4.8 hour modulation, thought to be the
orbital period of the binary, is observed at X-ray 
(\cite{brink}, \cite{pars}) and infrared 
(\cite{bec72}, \cite{mason}) wavelengths.  The 4.8 hour orbital 
period is typical of a low mass X-ray binary, but infrared observations 
suggest that the mass donating companion is a high mass 
Wolf-Rayet star (\cite{vank}). Due to the 
distance ($>$ 8.4 kpc; see \cite{boncha}), and its location in the Galactic 
plane, no optical counterpart has been found.

Cyg X-3 is also a strong source of radio emission with three types of behavior:  (a) quiescence, 
(b) major flaring 
with quenching (very low radio fluxes), 
and (c) minor radio flaring with partial 
quenching (see Waltman et al. 1994, 1995, 1996).
During the major radio outbursts 
there is strong evidence of 
jet-like structures moving away from Cyg X-3 at a velocity of either $\sim$ 0.3c or $\sim$ 0.6c, depending 
upon whether the jets are one-sided or double-sided (
\cite{moln}, \cite{schal}).  
Recent VLBA images
after a large Cyg X-3 flare showed one-sided ejection with v $\gtrsim$ 0.8c  (\cite{miod}).

     The soft X-ray emission (below 20 keV) from Cyg X-3 has been observed to 
undergo high and low states in which the various spectral components change 
with no apparent correlation with other properties (\cite{white2}).  
From Ginga observations (\cite{watan}), it was found that 
the giant radio flares occur only when the soft X-ray flux is high.
Hard X-ray (HXR) observations (above $\sim$ 50 keV),
prior to 1991, found the HXR 
emission to vary by less than 20\% (\cite{white1}, \cite{herm}).  Recent Compton Gamma-Ray Observatory (CGRO)
observations of Cyg X-3 made using OSSE have shown HXR emission to be variable by a factor of 3 
(\cite{matz}).    

In this letter, we report results obtained from the nearly continuous monitoring
of Cyg X-3 for a period of almost 5 years in both the HXR (CGRO/BATSE)
and the radio wavelength bands.  Cyg X-3 has been monitored in the HXR and radio as
part of a study to understand the high-energy properties of binary X-ray
sources and how they relate to the production of jets.

\section{HARD X-RAY OBSERVATIONS}

     The BATSE experiment onboard CGRO (\cite{fish}) was used to 
monitor
the HXR emission from Cyg X-3.  The BATSE Large Area Detectors (LADs) 
can monitor the whole sky almost 
continuously in the energy range of 20 keV--2 MeV with a typical daily 3 $\sigma$ 
sensitivity of about 100 mCrab.  
  
     To produce the Cyg X-3 light curve, single step occultation data were 
obtained using a standard Earth occultation analysis 
technique for  monitoring HXR sources (\cite{harm}).   
Interference from known bright sources (such as Cyg X-1) and single steps 
with large deviations from the median flux measurement (the result of the HXR
flaring behavior of Cyg X-1) were removed.  The single 
occultation step data were then 
fit with a power law with a fixed photon spectral index of $-3.0$ to determine a flux 
in the 20--100 keV band.  The spectral index was chosen based on spectral fitting of the BATSE data during
periods of high levels of emission and is consistent with the 
index found by \cite{matz} from OSSE observations.  
Varying the spectral index, for an acceptable range of spectral
indices, results in a 10--15 \% variation in the measured flux.  

A Lomb-Scargle periodogram (\cite{lomb}, \cite{scar}) was created for the single step occultation data to 
search for periodicities in the data.  The known 4.8 hour period of Cyg X-3 was detected
in the BATSE data set at a very high significance (\cite{rob}), with no additional periodicities being 
found in the data.  Sums of single step occultation data of $\geq$ 2 day duration were found to
uniformly sample all phases of the 4.8 hour period. 
 
\subsection{\bf Hard X-Ray Light Curve}
 
    Fig. 1 shows a history of the 20--100 keV HXR emission 
from Cyg X-3.   The BATSE data 
were binned with each point containing 40 occultation steps, with each occultation step having an
integration time of 110 seconds.  This corresponds to   
about three days of observations per data point.      
Cyg X-3 ranges from less than 
BATSE's three day detection limit of $\sim$~40 mCrab to intensities at least 7 times 
brighter.  The two most distinctive
features in the HXR light curves are (a) extended periods of high  
HXR flux (0.04 -- 0.08 ${\rm ph~cm^{-2}~sec^{-1}}$) and (b) low flux periods where the
flux level drops below the detection limit of BATSE.

\section{RADIO OBSERVATIONS}

    Cyg X-3 was observed on a daily basis as part of the Naval Research Laboratory 
Green Bank Interferometer (GBI) Monitoring 
Program.  The GBI
consists of two 25.9 m diameter antennas separated by a 2.4 km
baseline.  The antennas each observe simultaneously at 2.25 and 8.3 GHz with 
35 MHz of total
bandwidth at each frequency.  
The
observing and data reduction techniques are described in \cite{fied},   
\cite{walt2}, and \cite{walt3}.  The average integration time for
individual scans is approximately 10 minutes with up to 12 scans made on the
source per day.  Typical errors (1 $\sigma$) in the GBI data set are flux 
density dependent
$\colon$ 4 mJy (2 GHz) or 6 mJy (8 GHz) for fluxes less than 100
mJy; 15 mJy (2 GHz) or 50 mJy (8 GHz) for fluxes $\sim$ 1 Jy.  
In Fig. 1 the 2.25~GHz radio flux densities are shown overlaid on a log scale for
the same time period as the HXR measurements.

\section{THE HARD X-RAY -- RADIO RELATIONSHIPS}

What is striking to the eye is an overall anticorrelation between the radio 
and the HXR emission.  At the times
when the radio has become quiescent the HXR increases and goes
into a state of high emission.  With the 
occurrence of radio flares the HXR emission drops and during the major flaring activity the
behavior of the HXR and radio changes to a correlation (see Fig. 2).

To test fully the existence of these (anti)correlations we 
calculated correlation 
coefficients ({$\rm r_s$}) and probability functions of the Spearman rank-order.  
To assure that each pair of points is statistically independent,
rejection of some fraction of the data to eliminate any internal correlation is 
necessary.\footnote{Care must be taken because if the time series contain
internal correlations then a correlation test between the two time series can result in either a false or
enhanced correlation (\cite{walth}).} 

For each of the two time series we divided the data into two sets of observations 
consisting of alternating measurements from the original 
time series.  These series were tested and if an internal 
correlation was found, we
then rejected half of the data points and repeated the procedure with the remaining data.  We continued this
process until at least one of the two different time series showed no internal 
correlation,  then  
tested the matching data 
points in the time series
of the two wavelength bands for a correlation.  In all cases we consider
a correlation to exist if the probability of no correlation was $10^{-3}$ or less.  

Table 1 shows the results of correlation testing using 3 day averages of the data for different sets 
of the data.  For
each type of activity is given (a) the Spearman correlation coefficient, (b) probability of no correlation,
(c) initial size of the data set, (d) size of the data set after data 
rejection,  and (e) the wavelength band (R: Radio, X:
HXR, B: Both) in which there is no internal correlation.

{\it Entire Data Sets:} During times of high HXR flux the radio data show less
variability and as flaring occurs, the HXR starts to drop.  One can define a
variability index by taking the standard deviation ($\sigma$) in
the radio data, over the time averaged interval, and dividing it by the mean 
value
of the radio flux ($\langle$S$\rangle$) for the time averaged interval.  
Table 1 and Fig. 3 shows the existence of a strong 
anticorrelation between the HXR flux and the variability index of the radio flux.

{\it Quiescence Radio Activity:} For the periods of quiescent radio emission in our data sets we 
see that there is a very
strong anticorrelation between the HXR and the radio fluxes.  

{\it Major Flaring Activity:}   
For major flaring activity  
one can see there exists a strong correlation between the HXR and the radio fluxes (see Fig. 2).  
OSSE observations made during the first major radio flare activity period 
showed a factor of 3 drop in
the HXR as the radio became quenched prior to a major flare and later observations 
showed a higher flux level as Cyg X-3 returned to quiescence (\cite{matz}) in agreement with the BATSE data.    

{\it Minor Flaring 
Activity:} During the minor flaring  
there was no significant correlation found between the HXR and radio fluxes.

Fig. 4 is a radio versus HXR flux plot for the radio quiescent periods and the major radio flare
activity.  
In this plot we see two branches which describe the correlated behavior which we have
found.  There is a horizonal branch which describes the quiescent radio state and a vertical branch
which describes the major radio flaring.  It is of interest to note that where the two branches overlap one 
finds the minor radio flaring data points.

\section{DISCUSSION}

For the first time, we report correlated HXR and radio emission in Cyg X-3 (see Table 2 for a
summary).  
The high degree of anticorrelation, during radio quiescence, between the wavelength bands indicates that 
the same high energy 
particle production mechanism may be responsible for the particles which produce both the radio 
(synchrotron) and hard 
X-ray (Compton processes).  However the efficiency of inverse-Compton processes (\cite{shu}) in cooling 
the high energy electrons
required to produce the synchrotron radio emission make it unlikely the HXR and radio 
originate from the same emission region.  
These high energy particles may be the result of either the production of a 
``hot'' corona (\cite{white1}) or the presence of a low-level jet (\cite{fender}), during a 
period of low mass transfer in the system.

The major radio flaring activity represents a distinct change in the behavior of the system.  
The quenched radio emission 
preceding large to medium 
radio flares 
clearly corresponds to low HXR states in the BATSE data. 
Such episodes have been interpreted (\cite{fender}), based on the 
increase in soft X-rays (\cite{kita}), as an indication of a large
increase in mass flow within an accretion disk prior to a major radio flare.
The nature of this radio emission during the preflare episodes is not
understood, but possibly both the HXR and radio fluxes
are generated in a region (initially optically thick) near the central object.  It is apparent that the
quenched radio emission and the low HXR state are tied to the formation of the jet.

When a large radio flare occurs, there is a corresponding increase in the HXR flux.  
As the radio flare decays, the HXR also drops on a similar timescale.  The
decay in the HXR may be interrupted by subsequent flares or a transition to a radio quiescent state. 
The radio flare can be interpreted as a
consequence  of an ejection event from a bipolar or one-sided jet (\cite{schal1}, 
\cite{miod}) where the radio flare is due to the growth in solid angle and expansion of the
developing jets.  The HXR emission is most likely the result of inverse Compton 
processes, but the site of this emission is less certain.  Early observations of the galactic jet sources, 
GRS 1915+105 (\cite{foster}) and GRO J1655-40 (\cite{harm1}), which showed a correlation between the HXR 
and radio, gave rise to
models in which the HXR emission arose from the jet (\cite{lev}).  But later observations of these two
sources have shown HXR outbursts with which there was no accompanying radio jets 
(GRO J1655-40; \cite{zhang}) or there was an anticorrelation between the radio flare and the HXR 
flux (GRS 1915+105; \cite{harm2}).  
While there may be 
uncertainty on the site of the HXR emission it is clear that there is HXR response for all of
the major flares (and associated jets) in Cyg X-3.

During minor radio flaring periods there occur smaller ``mini-flares" and  a drop in the overall HXR flux 
which is reflected in the anti-correlation between HXR  
flux and the radio variability index.  
It is possible that conditions exist here similar to those present during the
major radio outbursts,
although the relatively low intensity in HXRs
makes it difficult to check if the faster variability seen in the radio
is also reflected at the same timescales in HXRs.

\cite{kita} have suggested that the X-ray and radio 
activity are mediated by the dense wind from the companion star.  
In the quiescent radio
state, the wind density is predicted to be low.  
Changes in the wind may
determine whether the source changes from a radio quiescence/HXR high state to radio flaring/HXR low state.

\section*{ACKNOWLEDGMENTS}

Basic research in radio astronomy at the Naval Research Laboratory is supported
by the Office of Naval Research.  The Green Bank Interferometer was operated by the National Radio Astronomy
Observatory under contract to the U.S. Naval Observatory during the time period of these observations.  
The National Radio Astronomy Observatory is operated by Associated Universities, Inc. under a
Cooperative Agreement with the National Science Foundation.

\begin{figure}                                                                         
\plotfiddle{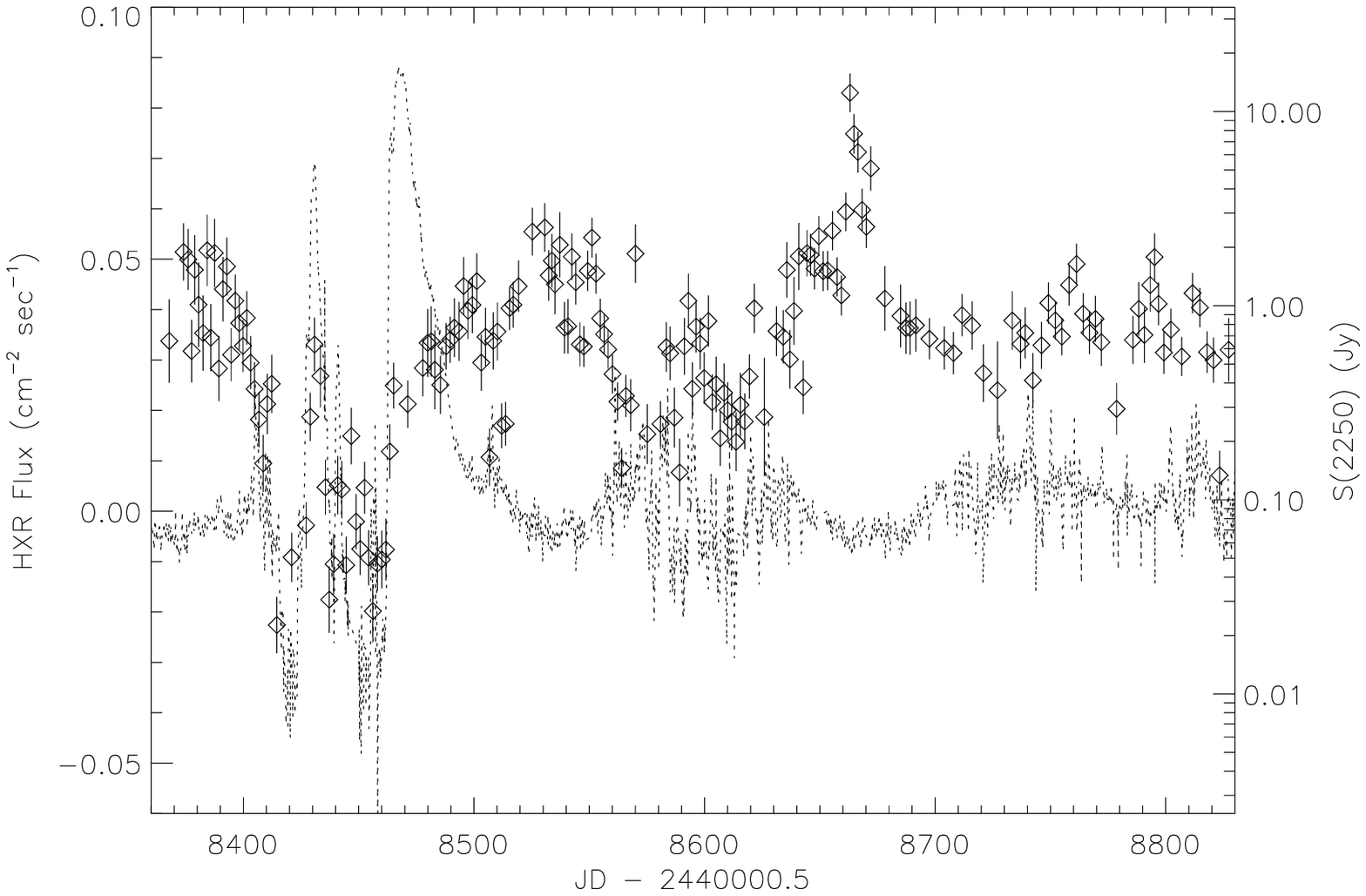}{1.35in}{0.0}{92.00}{37.00}{-288.0}{-120.0}
\plotfiddle{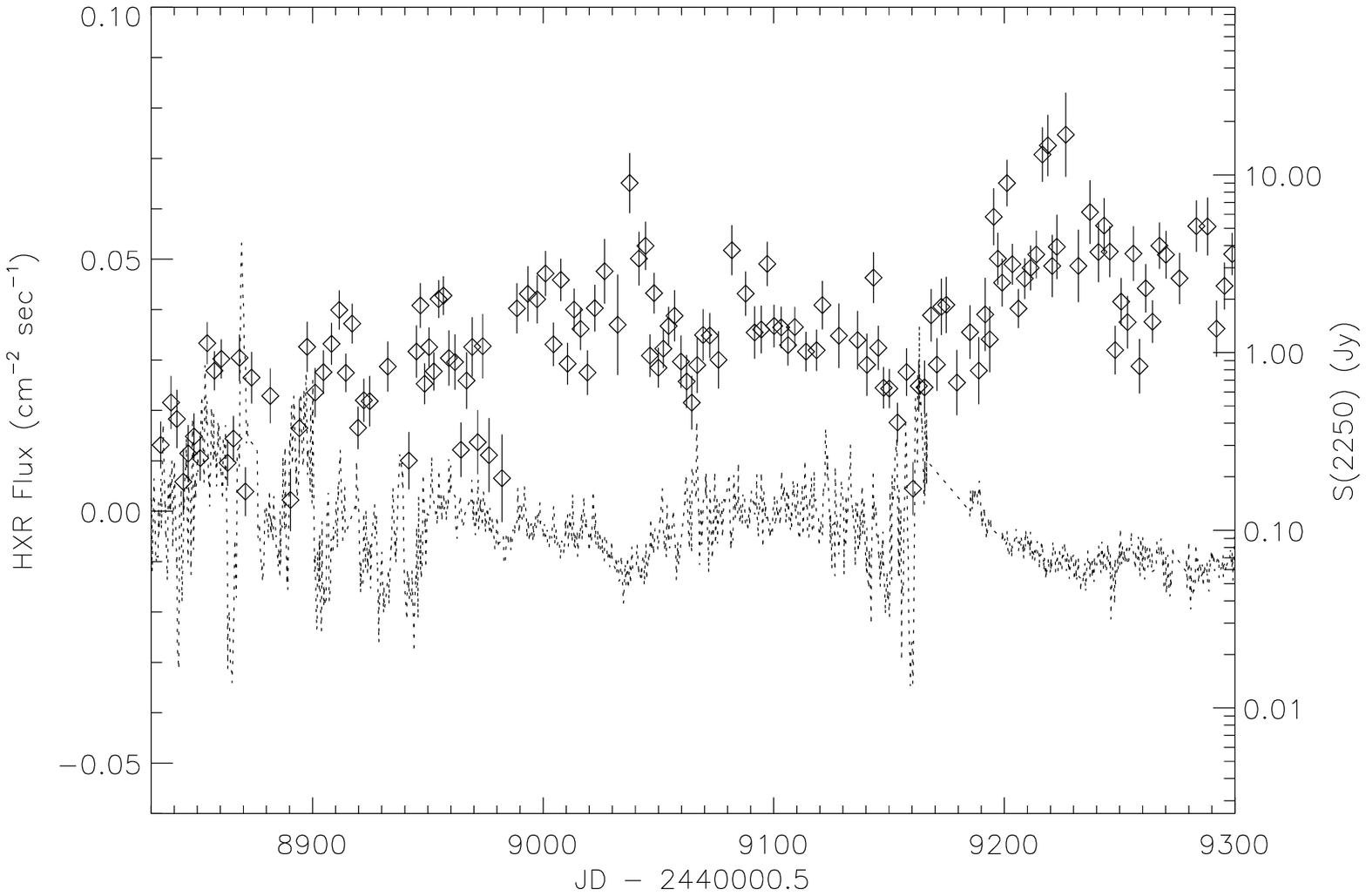}{1.35in}{0.0}{92.00}{37.00}{-288.0}{-128.0}
\plotfiddle{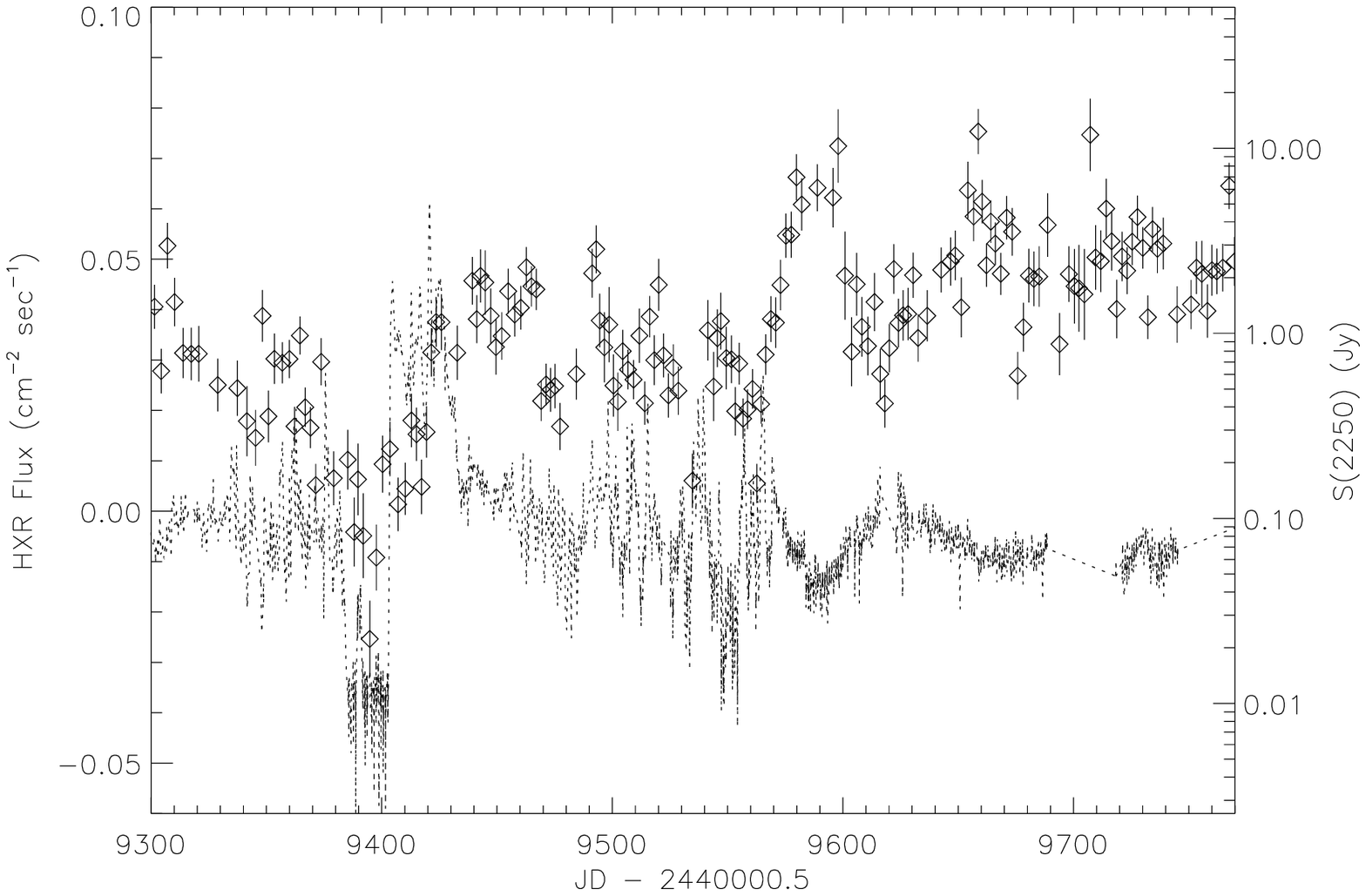}{1.35in}{0.0}{92.00}{37.00}{-288.0}{-136.0}
\plotfiddle{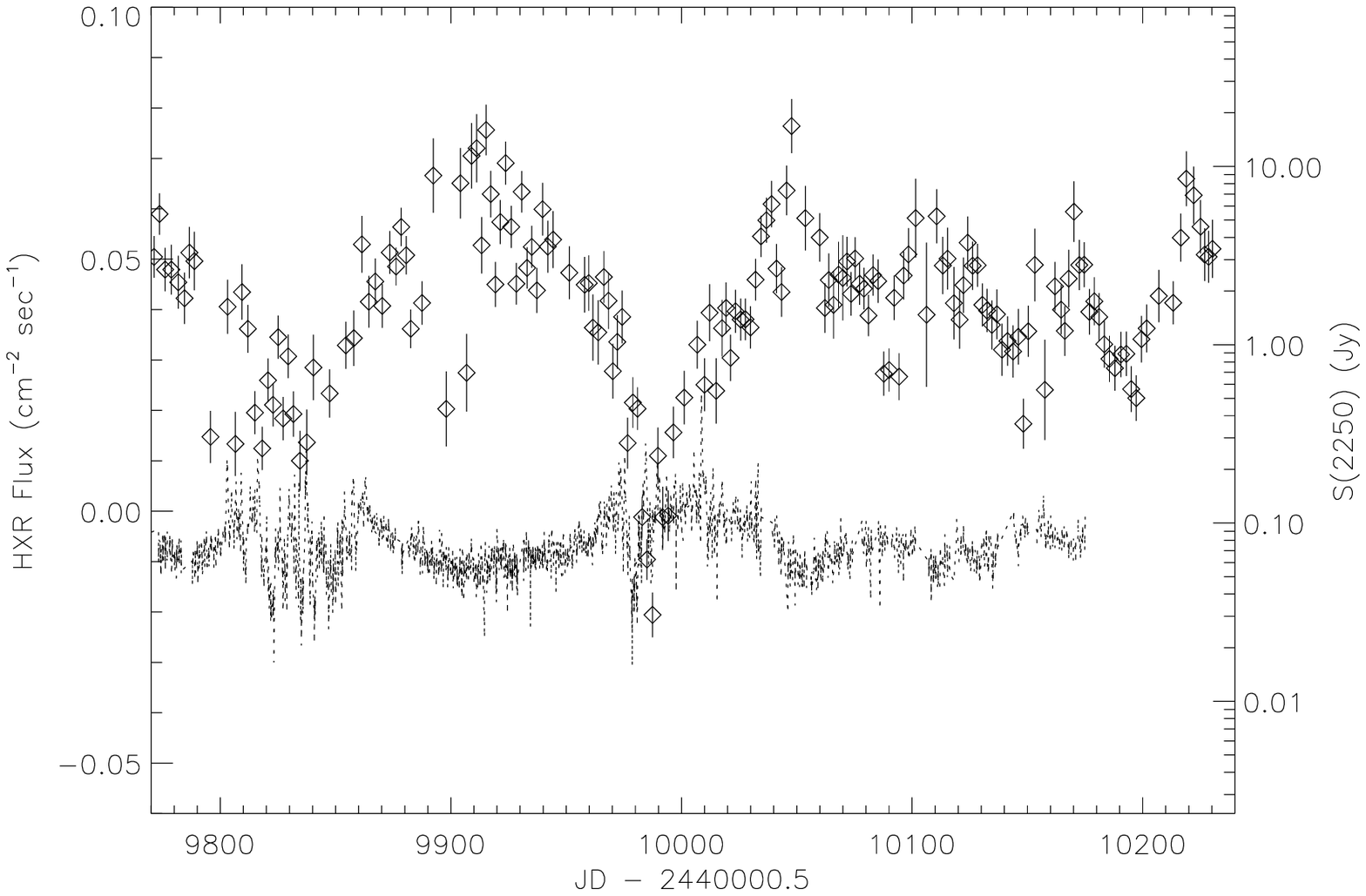}{1.35in}{0.0}{92.00}{37.00}{-288.0}{-144.0}
\caption{ A plot of the BATSE 20 - 100 keV flux of Cyg X-3 (diamonds with error 
bars).   
Overlayed on a log scale is the 2.25 GHz flux (dotted 
line) measured by the GBI during the same time interval.  The left hand scale (${\rm ph ~cm^{-2} sec^{-1}}$) 
is for the HXR flux and the right hand scale (Jy) is for the  2.25 GHz flux.}
\end{figure}

\begin{figure}                                                                  
\plotone{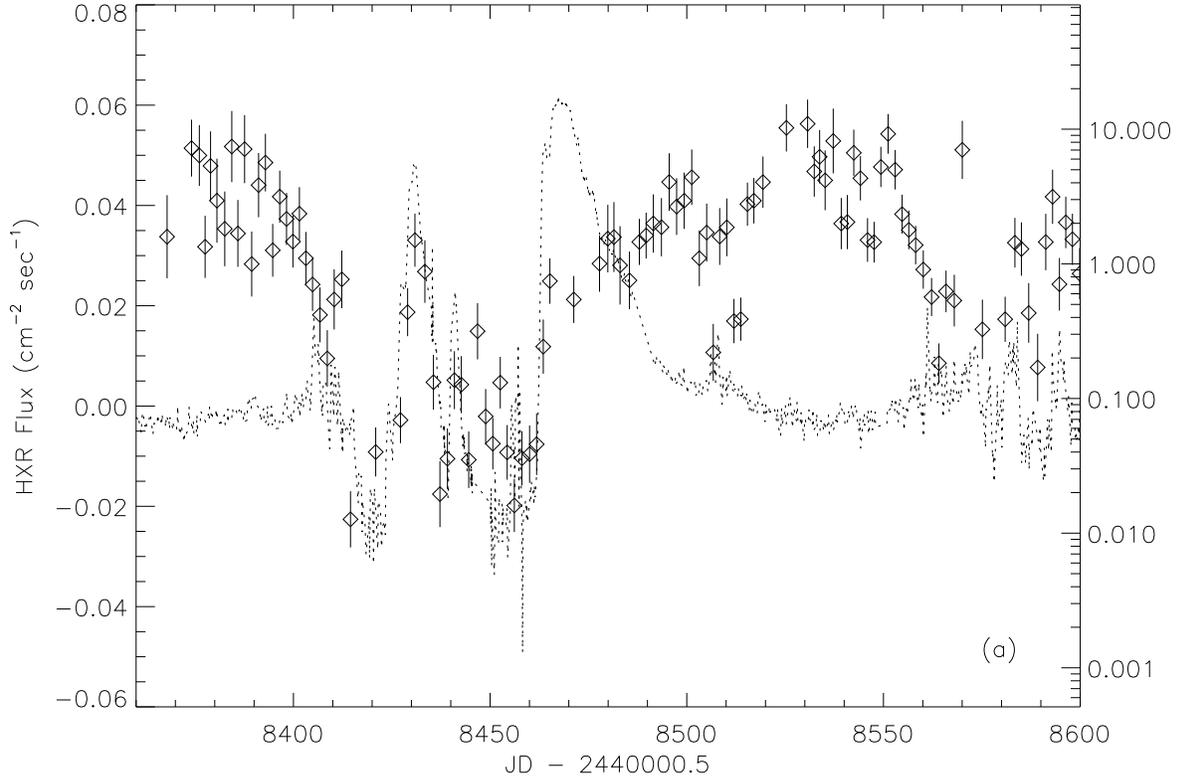}
\caption{HXR flux from Trucated Julian Dates (TJD) 8360 to 8600. Overlaid is the 2.25
GHz flux (dotted line). One can see the transition from a radio quiescent state (anticorrelated with the
HXR) into quenched emission and major flares (correlated with HXR) back into a radio quiescent state
(anticorrelated with the HXR).  One can also see the transition, in the latter radio quiescent state, into a
minor flaring state.}
\end{figure}

\begin{figure}                                                                  
\plotone{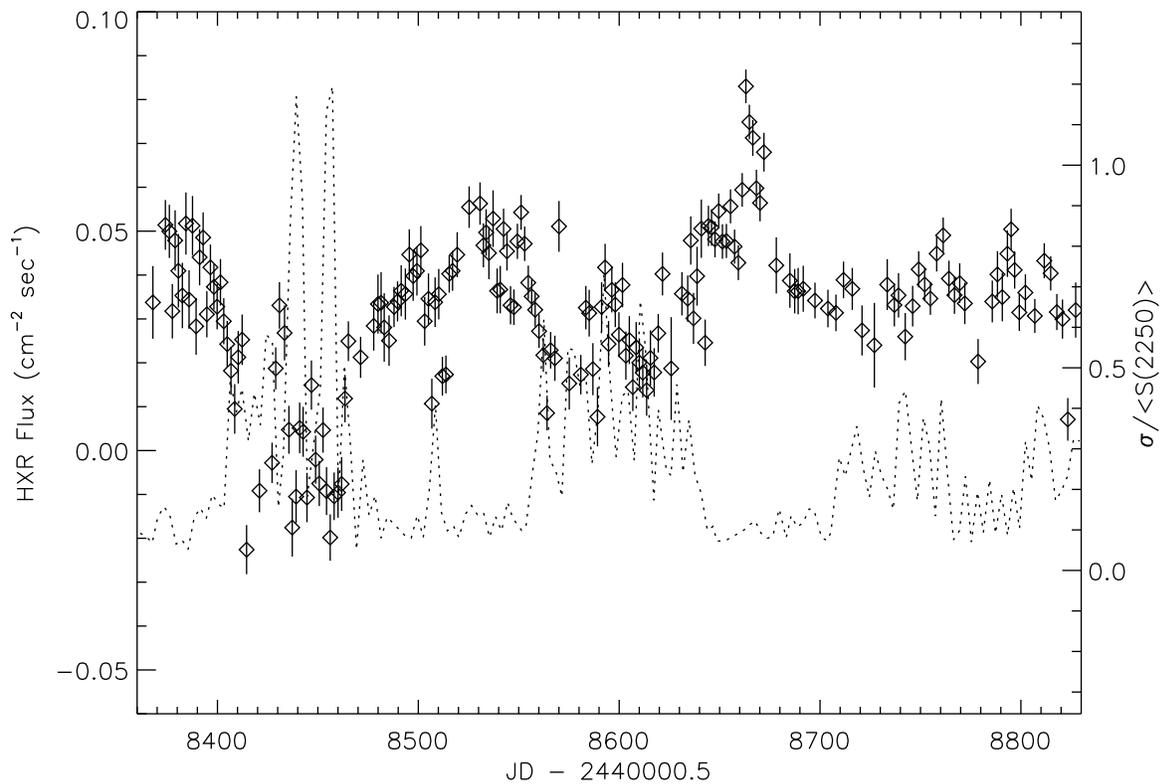}
\caption{HXR flux from TJD 8360 to 8830 (same time
range as the top panel of Fig. 1). Overlaid is the variability index of the 2.25
GHz flux (dotted line) for three-day averages.  The variability index is defined as 
$\sigma$/$\langle$S(2250)$\rangle$ where
$\sigma$ is the standard deviation of the data being averaged and 
$\langle$S(2250)$\rangle$ is
the mean flux of the radio data being averaged.}
\end{figure}

\begin{figure}                                                                  
\plotone{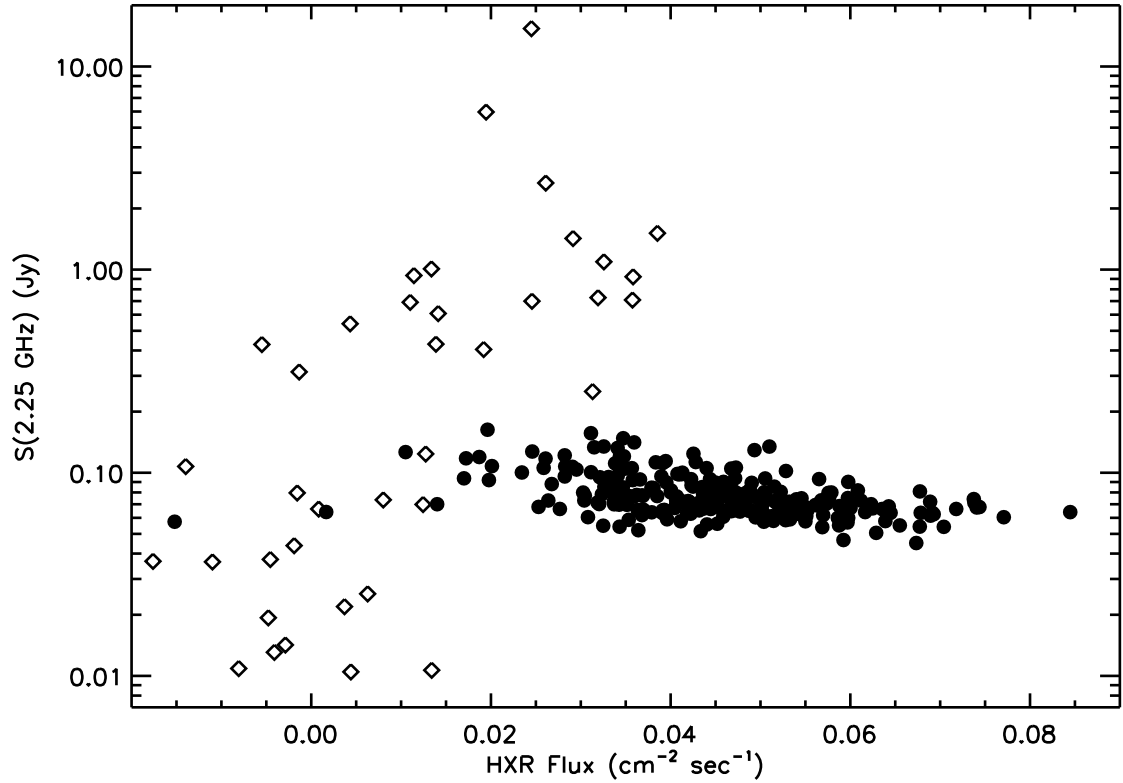}
\caption{This plot contains both the radio quiescent periods (filled circles) and the major flare activity
(diamonds) for three day averages. For clarity
the minor flaring data (which does not show a correlation) are not plotted, but you would find the
majority of minor flaring data points contained within a box defined by 0.0-0.045 
${\rm ph~cm^{-2}~sec^{-1}}$ in the HXR and 0.05--0.30 Jy in the radio.  This region corresponds to the area 
in which radio quiescent periods and the major flare activity overlap one another.}
\end{figure}

\begin{deluxetable}{lccccc}
\tablecolumns{6}
\small
\tablewidth{0pc}
\tablecaption{Tests of Correlation Between the Data Sets}
\tablehead{
\colhead{}   &  \multicolumn{5}{c}{Spearman}  \\
\cline{2-5} \cline{5-6} \\
\colhead{Source Activity} & \colhead{${\rm r_s}$} & \colhead{Prob.} & 
\colhead{Initial} & \colhead{Final} & \colhead{Band}}
\startdata
{\bf All Data:} \nl
HXR vs. $\sigma$/$\langle$S(2250)$\rangle$ & --0.489 & 2 x $10^{-9}$ & 546 & 136 & R \nl
HXR vs. $\sigma$/$\langle$S(8300)$\rangle$  & --0.405 & 1 x $10^{-6}$ & 546 & 136 & R \nl
{\bf Quiescence:} \nl
HXR vs. 2.25 GHz & --0.477 & 1 x $10^{-8}$ & 254 & 127 & X \nl
HXR vs. 8.3 GHz & --0.491 & 4 x $10^{-9}$ & 254 & 127 & X \nl
{\bf Flaring:} \nl
HXR vs. 2.25 GHz (Major) & 0.738 & 5 x $10^{-8}$ & 40 & 40 & B \nl
HXR vs. 8.3 GHz (Major) & 0.663 & 3 x $10^{-6}$ & 40 & 40 & B \nl
HXR vs. 2.25 GHz (Minor) & 0.168 & 9 x $10^{-3}$ & 243 & 243 & R \nl
HXR vs. 8.3 GHz (Minor) & 0.190 & 4 x $10^{-2}$ & 243 & 121 & B \nl
\enddata
\end{deluxetable}

\begin{deluxetable}{lll}
\tablecolumns{3}
\small
\tablewidth{0pc}
\tablecaption{Summary of HXR - Radio Relationship}
\tablehead{\colhead{Radio Activity} & \colhead{Relationship} & Time Periods (TJD)} 
\startdata
{\bf All} & There exists a anticorrelation between  & 8361--10175 \nl
& HXR and the variability in the radio. \nl
\nl
{\bf Quiescence} & The HXR and radio show a strong   & 8361--8400, 8500--8557, 8635--8705,  \nl 
& anticorrelation.  It is during this  & 8960--9050, 9170--9330, 9570--9610, \nl
& state that the HXR reaches its & 9630--9800, 9860--9975, 10038--10175 \nl
& highest levels. \nl
\nl
{\bf Major Radio} & The HXR and radio correlate with  & 8415--8500, 8860--8876, 9385--9432 \nl 
{\bf Flaring} & the HXR vanishing during quenching \nl
& and recovering during the major \nl
& flare.\nl
\nl
{\bf Minor Radio} & The HXR and radio show no & 8557--8635, 8705--8860, 8876--8960, \nl
{\bf Flaring} & correlation for this activity. & 9050--9170, 9330--9385, 9432--9570, \nl
& & 9610--9630, 9800--9860, 9975--10038 \nl
\enddata
\end{deluxetable}

\end{document}